\documentclass[prfluids,floats,showpacs,floatfix,superscriptaddress,longbibliography]{revtex4-1}
\usepackage{graphics,graphicx,dcolumn,bm,epic,eepic,float}
\usepackage{amssymb,amsmath,multirow,rotate,color,float,mciteplus}
\usepackage[latin1]{inputenc}
\usepackage{color}
\usepackage[dvips]{epsfig}

\usepackage{epstopdf}

\begin{document}

\title{Boundary conditions at the gas sectors of superhydrophobic grooves.}

\author{Alexander L. Dubov}
\affiliation{A.N.~Frumkin Institute of Physical
Chemistry and Electrochemistry, Russian Academy of Sciences, 31
Leninsky Prospect, 119071 Moscow, Russia}
\author{Tatiana V. Nizkaya}
\affiliation{A.N.~Frumkin Institute of Physical
Chemistry and Electrochemistry, Russian Academy of Sciences, 31
Leninsky Prospect, 119071 Moscow, Russia}
 \author{Evgeny S. Asmolov}
\affiliation{A.N.~Frumkin Institute of Physical
Chemistry and Electrochemistry, Russian Academy of Sciences, 31
Leninsky Prospect, 119071 Moscow, Russia}
\affiliation{Institute of Mechanics, M.V.~Lomonosov Moscow State
University, 119991 Moscow, Russia}
\author{Olga I. Vinogradova}
\email[Corresponding author: ]{oivinograd@yahoo.com}
\affiliation{A.N.~Frumkin Institute of Physical
Chemistry and Electrochemistry, Russian Academy of Sciences, 31
Leninsky Prospect, 119071 Moscow, Russia}
\affiliation{Department of Physics, M.V.~Lomonosov Moscow State University, 119991 Moscow, Russia}
\affiliation{DWI - Leibniz Institute for Interactive Materials, Forckenbeckstra\ss e 50, 52056 Aachen,
  Germany}

\date{\today}
\begin{abstract}

The hydrodynamics of liquid flowing past gas sectors of unidirectional superhydrophobic surfaces is revisited. Attention is focussed on the local slip boundary condition at the liquid-gas interface, which is equivalent to the effect of a gas cavity on liquid flow. The system is characterized by a large viscosity contrast between liquid and gas, $\mu/\mu_g \gg 1$. We interpret earlier results, namely the dependence of the local slip length on the flow direction, in terms of a tensorial local slip boundary condition and relate the eigenvalues of the local slip length tensor to the texture parameters, such as the width of the groove, $\delta$, and the local depth of the groove, $e(y, \alpha)$. The latter varies in the direction $y$, orthogonal to the orientation of stripes, and depends on the bevel angle of groove's edges, $\pi/2 - \alpha$, at the point, where three phases meet. Our theory demonstrates that when grooves are sufficiently deep their eigenvalues of the local slip length tensor depend only on $\mu/\mu_g$, $\delta$, and $\alpha$, but not on the depth. The eigenvalues of the local slip length of shallow grooves depend on $\mu/\mu_g$ and $e(y, \alpha)$, although the contribution of the bevel angle is moderate. In order to assess the validity of our theory we propose a novel approach to solve the two-phase hydrodynamic problem, which significantly facilitates and accelerates calculations compared to conventional numerical schemes. The numerical results show that our simple analytical description obtained for limiting cases of deep and shallow grooves remains valid for various unidirectional textures.

\end{abstract}


\maketitle

\section{Introduction}

Anisotropic superhydrophobic (SH) surfaces have raised a considerable interest over the recent years. Such surfaces in the Cassie state, i.e., where the texture is filled with gas, can induce exceptional lubricating properties~\cite{bocquet2007,rothstein.jp:2010,vinogradova.oi:2012} and generate secondary flows transverse to the direction of the applied pressure gradient~\cite{feuillebois.f:2010b,schmieschek.s:2012,zhou.j:2012}. These are important for a variety of applications that involve a manipulation of liquids at the small scale and can be used to separate particles~\cite{LabChip} and enhance their mixing rate~\cite{ou.j:2007,nizkaya.tv:2017}  in microfluidic devices.
During past decade the quantitative understanding of liquid flow past SH anisotropic surfaces was significantly
expanded. However, many fundamental issues still remain unresolved.

To quantify the drag reduction and transverse hydrodynamic phenomena
 associated with  SH surfaces with given area of gas and solid fractions it is convenient to construct the effective slip boundary condition applied at the imaginary homogeneous surface~\cite{vinogradova.oi:2011,Kamrin_etal:2010}, which mimics the real one and is generally a tensor~\cite{Bazant08}. Once eigenvalues of the slip-length tensor, which depend on the local slip lengths at the solid and gas areas, are determined, they can be used to solve various hydrodynamic problems. To calculate these eigenvalues, SH surface is usually modeled as a perfectly smooth with patterns of local boundary conditions at solid and gas sectors.  It is widely accepted that
one can safely impose no-slip at the solid area, i.e., neglect slippage of liquid past smooth solid hydrophobic areas,
which is justified provided the nanometric slip is small
compared to parameters of the texture~\cite{joly.l:2006,vinogradova:03,charlaix.e:2005,vinogradova.oi:2009}. For gas sectors of SH surfaces the situation is much less clear. Prior work often applied  shear-free boundary conditions at the flat gas areas~\cite{philip.jr:1972,priezjev.nv:2005,lauga.e:2003}. In this idealized description both a meniscus curvature and a viscous dissipation in the gas phase, which could affect the local slippage, are fully neglected.

Several groups have recently studied the effect of a meniscus curvature on the friction properties of SH surfaces~\cite{lauga2009,sbragaglia.m:2007,harting.j:2008,karatay.e:2013,crowdy.dg:2017}. Most of these studies neglected a viscous dissipation in the gas, by focussing on the connection of the meniscus protrusion angle and effective slip length (but note that no attempts have been made to connect the meniscus curvature and a local slip at the gas area). It has been generally concluded that if the protrusion angle is $\pm \pi/6$ or smaller the effective slip of the SH surface does not differ significantly from expected in the case of the flat interface, so that the model of a flat meniscus can always serve as a decent first-order approximation.

\begin{figure}[tb]
\includegraphics[width=0.75\columnwidth]{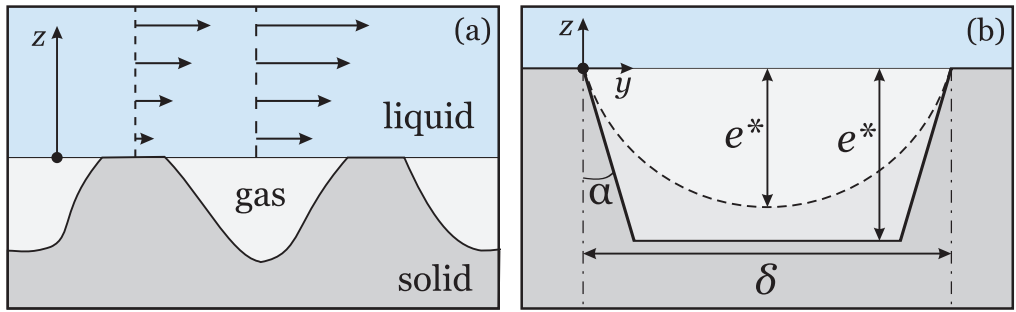}
\caption{(a)~Sketch of the flow past a 1D superhydrophobic surface. (b)~Sketch of trapezoidal (solid curve) and arc-shaped (dashed curve) grooves of width $\delta$, maximal depth $e^{\ast}$, and bevel angle $\pi/2-\alpha$.}
\label{fig_sketch}
\end{figure}

There is some literature describing attempts to provide a satisfactory theory of a local slip length, which take into account a viscous dissipation in the gas phase (and as far as we know, all these studies have modeled the liquid-gas interface as flat). We mention below what we believe are the most relevant contributions. To account for a dissipation in gas it is necessary to solve Stokes equations both for the liquid and for the gas phases (Fig.~\ref{fig_sketch}(a)), by imposing boundary conditions
\begin{equation}
 z=0: \; \mathbf{u}= \mathbf{u}_g, \; \mu \dfrac{\partial  \mathbf{u}_\tau}{\partial z}=\mu_g\dfrac{\partial \mathbf{u}_{g\tau}}{\partial z},
\label{BCcon}
\end{equation}
where $\mathbf{u}$ and $\mu $ are the velocity and the dynamic viscosity of the liquid, and $\mathbf{u}_g$ and $\mu _g$ are those of the gas, $\mathbf{u}_\tau=(u_x,u_y)$  is the tangential velocity.  This  problem has been resolved numerically for rectangular grooves~\cite{maynes2007,ng:2010}. It is however advantageous to  replace the two-phase approach, by a single-phase problem  with spatially dependent partial slip boundary conditions~\cite{cottin:2004,belyaev.av:2010a,bocquet2007}. For unidirectional (1D) surfaces they are normally imposed as
\begin{equation}
 z=0: \; \mathbf{u}_\tau-b(y)\dfrac{\partial \mathbf{u}_\tau}{\partial z}=0,
\label{BCslip}
\end{equation}
where $b(y)$ is the local scalar slip length, which is varying in one direction only.

That the gas flow can be indeed excluded from the analysis being equivalently replaced by $b(y)$ is by no means obvious. Early work has suggested that the effect of gas-filled cavities is equivalent to the introduction of a slip length, proportional to their depth if shallow and to their spacing if deep~\cite{hocking.lm:1976}. To model the trapped gas effects theoretically a semi-analytical method, which predicted a non-uniform local slip length distribution across the liquid-gas interface, has been proposed~\cite{shoenecker.c:2014,shoenecker.c:2013}. During a last few years several authors have  discussed that the local slip depends on the flow direction~\cite{shoenecker.c:2013,nizkaya.tv:2013,nizkaya2014gas}. Although they did not fully recognize it, their results are equivalent to a tensorial generalization of  Eq.(\ref{BCslip})

\begin{equation}
 z=0: \; \mathbf{u}_\tau-\mathbf{b}\dfrac{\partial \mathbf{u}_\tau}{\partial z}=0,
\label{BCslip_tensor}
\end{equation}
where $\mathbf{b}=b\{i,j\}$ is the second-rank local slip length tensor, which is  represented  by
symmetric,  positive definite  $2\times 2$ matrix diagonalized by a rotation with respect to the alignment of the SH grooves.

Recent work has elucidated  a mechanism which transplants the flow in the gas to a local slip boundary condition at the liquid-gas interface~\cite{nizkaya2014gas}. This study has concluded that the  non-uniform local slip length of a shallow texture is defined by the viscosity contrast and local thickness of a gas cavity, similarly to infinite systems~\cite{vinogradova.oi:1995a,miksis.mj:1994}. In contrast, the local slip length of a deep texture has been shown to be fully controlled by the dissipation at the edge of the groove, {i.e.} the point where three phases meet, but not by the texture depth as has sometimes been invoked for explaining the extreme local slip.
These results led to simple formulas describing liquid slippage at the trapped gas interface of 1D grooves of width $\delta$ and constant depth $e^*$~\cite{nizkaya2014gas}

\begin{equation}\label{bgas}
b_{\|,\perp} \simeq \dfrac{\mu }{\mu_g} \delta \beta_{\|,\perp},
\end{equation}
where  $b_{\|,\perp}$  are eigenvalues of the slip length tensor, $\mathbf{b}$, and $\beta_{\|,\perp}$ are eigenvalues of the tensorial slip coefficient, $\boldsymbol{\beta }$. The latter become linear in $e^*/\delta$, when $e^*/\delta$ is small and saturate at large $e^*/\delta$.
The validity of this ansatz for rectangular grooves has been confirmed numerically~\cite{nizkaya2014gas} and, although indirectly, experimentally~\cite{nizkaya.tv:2016}.

Previous investigations  have addressed the question of local slip at the gas areas of rectangular grooves with a constant depth only. We are unaware of any previous work that has quantified liquid slippage at gas areas of more general 1D SH surfaces. In this paper, we explore grooves with beveled edges and a non-uniform depth, which varies with $y$ and depends on the bevel angle, $\pi/2 - \alpha$ (see Fig.~\ref{fig_sketch}(b)). An obvious practical advantage of such a relief is that the manufactured grooves become mechanically more stable against bending compared to rectangular ones. This is especially important for dilute textures of large $\delta$, which induce larger slip. It is therefore very timely to understand important consequences of a bevel angle for a generation of liquid slippage at the gas areas of such grooves. Here we present theoretical arguments, which allow one to relate $b_{\|,\perp}$ to $e(y, \alpha)$. Our results show that
$b_{\|,\perp}$ are not really sensitive to a bevel angle when SH grooves are shallow and weakly slipping, but the large local slip at deep SH grooves is controlled  by their width and bevel angle only. These two parameters could be used to tune the large slip at the gas areas of any grooved SH surface and constrain its attainable upper value.

 Our paper is organized as follows. In Sec.~II we describe our model and justify the choice of model surfaces. Sec.~III gives a brief summary of our theoretical results for $b_{\|,\perp}$ obtained in some limiting situations. In Sec.~IV we describe a numerical method developed here to compute local slip length profiles. To solve numerically the two-phase hydrodynamic problem we consider separately flows in liquid and gas phases, which is in turn done by using different computational techniques. Our results are discussed in Sec.~V, and we conclude in Sec.~VI. Details of our asymptotic analysis can be found in Appendix~\ref{sec_app_edge}.

\section{Model}

We consider  creeping flow past 1D SH textures of period $L$  and gas area fraction $\phi$. The coordinate axis $x$ is
parallel to the grooves; the cross-plane coordinates are denoted by $y$ and $z$. The width of the groove is denoted as $\delta=L \phi$. The bevel angle of the grooves is $\pi/2 - \alpha$, where an angle $\alpha\leq \pi/2$ is defined relative to the vertical, and the depth of the grooves, $e(y, \alpha)$, is varying in only one direction  (see Fig.~\ref{fig_sketch}(b)). Since we consider SH textures with air
trapped in the grooves in a contact with water, the  ratio of liquid and gas viscosities is typically $\mu / \mu_g \simeq 50$, which is much larger than unity. Our results apply to a situation where the capillary and Reynolds numbers are sufficiently small, so that the liquid-gas interface does not deform, but not to the opposite case, where significant deformation of this interface is expected~\cite{seo.j:2015}.

 \begin{figure}[tb]
\includegraphics[width=0.375\columnwidth]{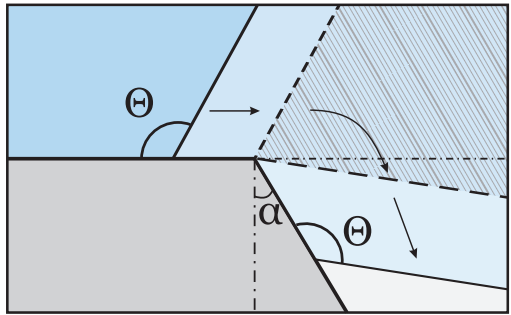}
\caption{Sketch of the contact angle, ``measured'' taking the horizontal as a reference, before, upon, and after the
corner.}
\label{fig_cassie}
\end{figure}

The texture material is characterized by the (unique) Young angle $\Theta$ (above $\pi/2$) measured taking the horizontal as a reference. Displacing a contact line to the sharp edge of a bevel angle $\pi/2 - \alpha$ we would observe an apparent variability of the contact angle from $\Theta$ to $\pi/2 - \alpha + \Theta$ since the line becomes pinned (see  Fig.~\ref{fig_cassie}). Using simple geometry one can then conclude that in the Cassie state  permitted protrusion angles of a meniscus are confined between $\pi/2 + \alpha - \Theta$ and $\pi - \Theta$, so that one of these (many) possible  protrusion angles, which would be observed in practice, should be determined by a pressure drop between the liquid and the gas phases. Since  positive protrusion angles (convex meniscus) are expected only when an external pressure is applied to the gas phase~\cite{karatay.e:2013}, we can exclude this artificial for SH surfaces case from our consideration. Now simple estimates suggest that for  $\Theta = 2 \pi /3$, typical for SH texture materials, and when $\alpha = 0$, bounds, which constrain attainable protrusion angles are $0$ (flat meniscus) and $- \pi /6$ (concave meniscus). Tighter bounds for finite $\alpha$ further constrain the attainable protrusion angle. These angles are too small to significantly reduce the effective slip~\cite{lauga2009,karatay.e:2013}. Therefore, to highlight the effect of viscous dissipation on the slippage in simpler terms we here assume that liquid-gas interface to be flat, which implies that pressure in gas is equal to that in liquid. Such a situation occurs when trapped by texture gas is in contact with the atmosphere~\cite{nizkaya.tv:2016}.

Our aim is to investigate how the local depth $e(y,\alpha)/\delta$ modifies the slippage of liquid past gas areas. We will provide some general theoretical arguments and results valid for any shape of the SH grooves, and also some simulation results for representative SH grooves. As an initial illustration of our theoretical and computational approach we will consider a  trapezoidal surface, where $\alpha$ and the maximal depth, $e^*$, can be varied independently and in the very large range
\begin{equation}
  \label{eq:profile}
  e(y, \alpha) =
  \begin{cases}
    y \cot \alpha, & y \le e^* \tan \alpha \\
    e^*, & e^* \tan \alpha < y \le \delta - e^* \tan \alpha \\
    (\delta - y) \cot \alpha, & \delta - e^* \tan \alpha < y
  \end{cases}
\end{equation}
We should like to mention that motivated by a recent experiment~\cite{choi.ch:2006} there have been already attempts to determine an effective slip past trapezoidal SH grooves~\cite{Zhou_etal:2013}, but that work has simply assumed the trapezoidal shape of a local scalar slip and no attempt has been made to properly connect it with viscous dissipation in the confined gas.

Another type of SH surfaces we explore here are grooves bounded by arcs of circles of radii $\displaystyle \delta/(2 \cos \alpha)$. These arc-shaped grooves do not have areas of a constant depth, and the $e(y, \alpha)$ profile does not contain sectors where dependence on $\alpha$ and $\delta$ disappears
\begin{equation}
e(y, \alpha)=\frac{\delta}{2}\left(\sqrt{\frac{1}{\cos^2\alpha}-\left(\frac{2y}{\delta}-1\right)^2}-\tan\alpha\right).
\label{eq:arc_groove}
\end{equation}
Their maximal depth, $e^{\ast}$, is attained at the midplane, $y/\delta=0.5$, and is given by
\begin{equation}\label{earc}
    e^{\ast} = \frac{\delta (1 - \sin \alpha)}{2 \cos \alpha}
\end{equation}
Eq.(\ref{earc}) shows that arc-shaped grooves can never be really deep since $ e^{\ast}/\delta$ cannot exceed $0.5$. It will be therefore instructive to compare their local slip with that of trapezoidal textures of the same $\delta$ and $e^{\ast}$.

\section{Theory}

Our aim is to calculate the eigenvalues of the local slip length tensor, $\mathbf{b}$, which can be found from the solution of the two-phase problem for the longitudinal  (fastest) and
transverse  (slowest) flow directions by imposing conditions, Eq.(\ref{BCcon})
\begin{equation}
b_{\parallel}=\left.\dfrac{u_{x}}{\partial u_{x}/ \partial z}\right|_{z=0} \equiv \dfrac{\mu }{\mu_g}\left.\dfrac{u_{g,x}}{\partial u_{g,x}/ \partial z}\right|_{z=0},
\label{bpadef}
\end{equation}
\begin{equation}
b_{\perp}=\left.\dfrac{u_{y}}{\partial u_{y}/ \partial z}\right|_{z=0} \equiv \dfrac{\mu }{\mu_g}\left.\dfrac{u_{g,y}}{\delta \partial u_{g,y}/ \partial z}\right|_{z=0}.
\label{bpedef}
\end{equation}
 However, since all properties of $\boldsymbol{\beta }$ are inherited by the local slip-length tensor, $\mathbf{b}$, below we will focus more on calculations of scalar non-uniform eigenvalues, $\beta_{\parallel,\perp}$~\cite{note3}.
In the general case they can be calculated only numerically. However, for some limiting cases explicit expressions can  be obtained.


The solution for local slip lengths can be found analytically close to the edge of the grooves, $y/\delta\ll 1$, and the details of our analysis are given in  Appendix~\ref{sec_app_edge}. Here we highlight only the main results.  Our theory predicts that
in the vicinity of groove edges, the eigenvalues of local slip length of any 1D texture augment from zero as

\begin{equation}
b_{\parallel} \simeq \frac{\mu }{\mu_g}\frac{2y}{\tan(\pi/4+\alpha/2)},\hspace{1cm} b_{\perp} \simeq \frac{b_{\parallel}}{4}.
\label{bxAsympt0}
\end{equation}%
In what follows that at small $y/\delta$ the slip coefficient grows as

\begin{equation}
\beta_{\parallel,\perp}\simeq \beta'_{\parallel,\perp} \dfrac{y}{\delta},
\label{angles}
\end{equation}
by
having slopes
\begin{equation}
\beta'_{\parallel} \simeq \dfrac{2}{\tan(\pi/4+\alpha/2)},\hspace{1cm}
\beta'_{\perp} \simeq \dfrac{1}{2\tan(\pi/4+\alpha/2)}.
\label{angles_prim}
\end{equation}
Note that when $\alpha = 0$, Eqs.(\ref{angles_prim}) predict $\beta'_{\parallel} \simeq 2$ and $\beta'_{\perp}  \simeq 1/2$. If $\alpha = \pi/2$, i.e. there are no gas sectors, $\beta'_{\parallel,\perp}  \simeq 0$.
We also remark that the linearity of the Stokes equations implies that near the second edge of the groove, $y/\delta = 1$, we have
\begin{equation}\label{angles2}
    \beta_{\parallel,\perp}\simeq \beta'_{\parallel,\perp}  \left(1 - \dfrac{y}{\delta}\right).
\end{equation}

It has been suggested before that local slip length of a shallow groove is defined only by the viscosity contrast and local thickness of a thin lubricating films~\cite{nizkaya.tv:2013}, similarly to infinite systems~\cite{vinogradova.oi:1995a,miksis.mj:1994}, but also depends strongly on the flow direction. Since in our case the local thickness depends on $\alpha$ the early results~\cite{nizkaya.tv:2013} can be generalized as
\begin{equation}
b_{\parallel}\simeq \dfrac{\mu }{\mu_g} e(y, \alpha),\hspace{1cm} b_{\perp}\simeq \frac{b_{\parallel}}{4},
\label{shallow0}
\end{equation}
and by using Eq.(\ref{bgas}) we can immediately formulate equations for eigenvalues of a slip coefficient in this limit
\begin{equation}
\beta_{\parallel}\simeq e(y, \alpha)/\delta,\hspace{1cm}\beta_{\perp}\simeq \beta_{\parallel}/4.
\label{shallow}
\end{equation}

Although Eqs.(\ref{angles}) and (\ref{angles2}) are valid for any groove depth, in the case of weakly slipping shallow groves the contribution of dissipation in the area $y/\delta\ll 1$ should be small compared to the central part of the gas areas, so that $\beta_{\parallel,\perp}$  are controlled by the groove depth as predicted by Eq.(\ref{shallow}). However, for strongly slipping deep grooves $\beta_{\parallel,\perp}$ should depend mainly on $\alpha$ since their maximal values are constrained by $\beta'_{\parallel,\perp}/2$. \label{refe1}This suggests that when grooves are deep, $\beta_{\parallel,\perp}$ are not sensitive to the shape of their bottom and should saturate at some $e(y, \alpha)/\delta$. Coming back to Eq.(\ref{bgas}) we conclude that local slip length at the gas areas of sufficiently deep grooves is defined by the viscosity contrast, their width, and bevel angle, but not by their depth. This implies that $\delta$ and $\alpha$ are the only two parameters that could be used to tune the large local slip at SH surfaces. An important remark would be that they also constrain its attainable upper value, and we can conclude that for rectangular grooves $b_{\parallel}$ should be inevitably below $\mu \delta/\mu_g$, and that $b_{\perp}$ is always smaller than $ \mu \delta/4\mu_g$. For grooves of the same $\delta$ with beveled edges these upper (and in fact unattainable) bounds will be even smaller.


\section{Computation of the two-phase flow near superhydrophobic grooves.}

A precise discussion of the flow in the vicinity of a SH surface requires a numerical solution of the self-consistent two-phase boundary problem, which is normally done by using finite element~\cite{girault2012} or boundary element~\cite{pozrikidis1992} methods. Here we suggest a simple alternative approach, which is easier to implement. We start by noting that Eq.(\ref{bgas}) does not contain any parameters associated with the flow of liquid. This implies that if Eq.(\ref{bgas}) is correct, then $\beta_{\parallel,\perp}$ are universal characteristics of the groove geometry only. However, applying the same $\beta_{\parallel,\perp}$ as boundary conditions yields different velocities at the gas-liquid interface, $u^{int}_{x,y}(y)=u_{x,y}|_{z=0}$, since the solution of the Stokes equations in liquid depends on  $\mu/\mu_g$ and $\phi$. This suggests that one can construct a simple iterative scheme to solve two-phase problem. We apply a lattice-Boltzmann method (LBM)~\cite{benzi1992,Luo1998} to simulate gas flow. It is robust, easy to implement, and allows one to measure velocity $\mathbf{u}_{g,\tau}$ and shear rate $\partial \mathbf{u}_{g,\tau}/ \partial z$ at the interface independently.

We  begin with a solution for $u^{int}_{x,y}$ obtained for an isolated perfect slip stripe, i.e. in the limit of small $\phi$~\cite{philip.jr:1972}
\begin{equation}
u_{x,y}^{int}/u^{\rm{max}}_{x,y} = \sqrt{1-(2y/\delta-1)^2},
\label{velprofile}
\end{equation}
and impose it as a boundary condition to calculate a flow in the gas phase. The eigenvalues of the local slip coefficient can then be calculated using Eqs.(\ref{bpadef}) and  (\ref{bpedef}). Note that it is also convenient to use a more accurate formula $\displaystyle u_{x,y}^{int}/u_{x,y}^{\max }=\frac{\cosh ^{-1}\left[ \cos \left( \pi
y\right) /\cos \left( \pi \phi /2\right) \right] }{\cosh ^{-1}\left[ 1/\cos
\left( \pi \phi /2\right) \right] }$, which is valid for an arbitrary $\phi$~\cite{philip.jr:1972,lauga.e:2003}, and that this velocity profile is very close to given by Eq.(\ref{velprofile}) when $\phi \leq 0.5$.

These computed eigenvalues specify   boundary conditions for liquid flow. We then solve Stokes equations in liquid numerically by using a method based on Fourier series~\cite{asmolov_etal:2013,nizkaya.tv:2013}. The solution satisfying these new partial slip conditions leads in turn to a better approximation for the velocity at the gas-liquid interface. At the next iteration we use this new velocity profile instead of Eq.(\ref{velprofile}) to compute more accurate values of $\partial \mathbf{u}_{g,\tau}/ \partial z$ and $\beta_{\parallel,\perp}$, and so on, until an exact solution is obtained.

We use D3Q19 implementation of LBM with the unit length set by the lattice step $a$, the time-step $\Delta t$, and mass density $\rho_0=1$. The kinematic viscosity of the fluid is  defined through the relaxation time scale $\tau$ as $\nu=(2\tau-1)/6$ and kept constant in all the simulations $\nu=0.03$. The simulated system is built in such way that the upper boundary represents a liquid-gas interface ($z=0$ in Fig.~\ref{fig_sketch}(b)) and the groove walls are perpendicular to $yz$-plane. For each simulation run the height of the system $N_z$ is chosen equal to the maximal depth of the groove. To provide sufficient resolution of groove shapes we use a simulation domain of the size $N_y = \delta = 126 a$, $N_x = 8 a$, and $N_z = 12 a - 122 a$. To verify the results we have repeated separate runs  with 2 times and 4 times larger space resolution and have shown that the maximal error due to discretization does not exceed~$1\%$. At the liquid-gas interface we set $\mathbf{u}_g|_{z=0}=(u^{int}_x,0,0)$ for longitudinal and $\mathbf{u}_g|_{z=0}=(0,u^{int}_y,0)$ for transverse grooves with $u_{x,y}^{int}/u^{\rm{max}}_{x,y}$ given by Eq.(\ref{velprofile}). The maximum velocity $u^{\rm{max}}_{x,y}$ at $y=\delta/2$ is taken equal to $10^{-3} a/\Delta t$. At the gas-solid interface, given by $z=-e(y)$, where $e(y)$  is defined by Eq.(\ref{eq:profile}) or Eq.(\ref{eq:arc_groove}), we impose no-slip boundary conditions.
For trapezoidal grooves we have varied $N_z/N_y=e/\delta$ from 0.1 to 0.97 and $\alpha$ from $0$ to $\pi/6$. For arc-shaped grooves the same values of $\alpha$ define $N_z=e^{\ast}$ (see Eq.(\ref{earc})). Finally, the periodic boundary conditions have been set along the $x$-axis. All simulations are made with an open-source package ESPResSo~\cite{ESPResSo}.

\begin{figure}[tb]
\includegraphics[width=0.75\columnwidth]{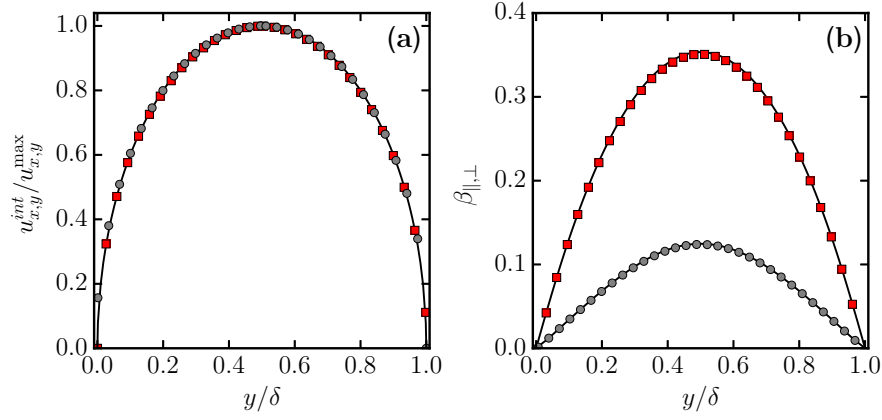}
\caption{(a) Dimensionless interface velocities $u_{x,y}^{int}/u^{\rm{max}}_{x,y}$ and (b)~eigenvalues of local slip coefficient $\beta_{\parallel,\perp}$ computed for a flow past trapezoidal grooves of $\alpha=3\pi/32$ and $\phi=0.5$.  Symbols plot the results of the second iteration obtained for longitudinal (squares) and transverse (circles) grooves. Solid curves correspond to the first iteration step.}
\label{fig_iterations}
\end{figure}

To  assess the validity and convergence of the approach  we  compute the interface velocity, $u_{x,y}^{int}/u^{\rm{max}}_{x,y},$ and the local slip coefficients, $\beta_{\parallel,\perp}$, at fixed  $\phi=0.5$, $\alpha=3 \pi/32$, and $e^*/\delta=0.97 $. Fig.~\ref{fig_iterations} shows the interface velocity profiles and $\beta_{\parallel,\perp}$ obtained after two iteration steps. We conclude that both longitudinal and transverse velocity at the gas sector nearly coincide with predictions of  Eq.(\ref{velprofile}) taken as an initial approximation, but we remark and stress that at the second iteration step $u^{\rm{max}}_{x,y}$ becomes quite different from used as an initial guess.  We also see that $\beta_{\parallel,\perp}$ computed at the first and the second iteration steps are practically the same. Therefore, in reality our iteration procedure converges extremely fast, so that below we use two iteration steps only.

Finally, to calculate the effective slip lengths for given computed local slip length profiles
\begin{equation}
 b_{\mathrm{eff}}^{\|,\perp}=\left.\dfrac{\langle u_{x,y}\rangle}{\langle \partial_z u_{x,y}\rangle}\right|_{z=0}
\end{equation}
we solve the Stokes equation in liquid numerically by using the Fourier series method described before in~\cite{nizkaya.tv:2013}. In these calculations we vary $\phi$ in the range from 0.1 to 0.9.

\section{Results and discussion}

\begin{figure}[tb]
\includegraphics[width=0.75\columnwidth]{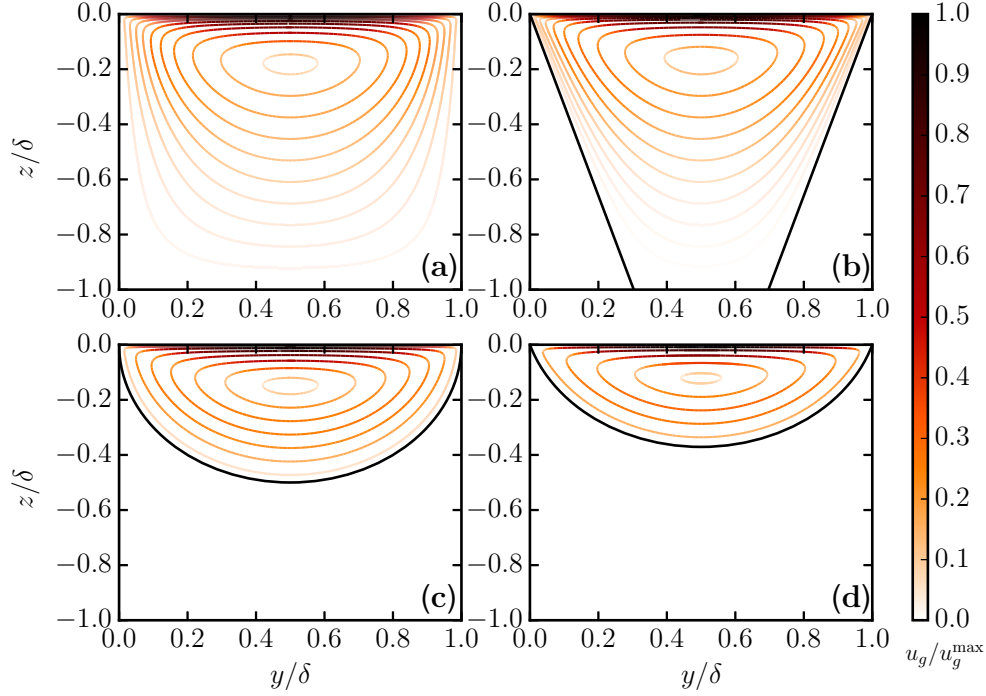}
\caption{Streamlines of the gas flow in (a,b)~the trapezoidal and (c,d)~arc-shaped transverse grooves of (a,c)~$\alpha=0$ and (b,d)~$\alpha=3\pi/32$. Color indicates the magnitude of velocity. Black lines show gas-solid interface.  }
\label{fig_strlines}
\end{figure}

We begin by studying the flow field in the gas phase of the grooves. For a longitudinal configuration the gas flows in $x$-direction only, and its velocity monotonously decays from gas-liquid interface to the bottom of the groove. The streamlines for a transverse case form a single eddy, and the locus of its center depends slightly on the shape of the groove and on $\alpha$. Fig.~\ref{fig_strlines} shows typical streamlines for a transverse flow in gas computed for trapezoidal grooves and arc-shaped grooves of $\alpha=0$ and $\alpha=3\pi/32$. The other parameters of a trapezoidal relief are taken the same as used in Fig.~\ref{fig_iterations}. Note that with these parameters the trapezoidal grooves are roughly twice deeper than arc-shaped. Thus, if $\alpha=0$, Eq.(\ref{earc}) gives $e^{\ast}/\delta = 1/2$ (cf. $e^*/\delta = 0.97$). Altogether the simulation results show that the velocity field in the gas phase is not a unique function of $\alpha$ and that it generally depends on the relief of grooves and their depth. We note, however, that when the grooves are sufficiently deep, the gas in them is almost stagnant near a bottom, so that increasing the depth further should not change the liquid flow past the SH surface.

\begin{figure}[tb]
\includegraphics[width=0.75\columnwidth]{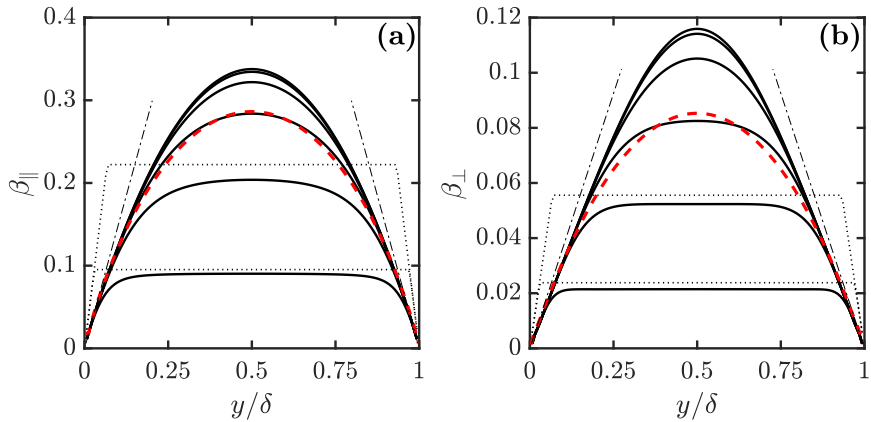}
\caption{(a) Longitudinal and (b)~transverse slip coefficients for trapezoidal grooves computed at $\alpha=3\pi/32$ (solid curves). From bottom to top $e^*/\delta=0.094$, $0.22$, $0.35$, $0.47$, $0.60$, and $0.97$. Dashed curves shows slip coefficient profiles for an arc-shaped groove with the same $\alpha$ and $e^{\ast}/\delta \simeq 0.37$. Dotted lines represent calculations with Eqs.(\ref{shallow}) made with $e^*/\delta=0.094$ and $0.22$. Dash-dotted lines are predictions of Eqs.(\ref{angles}) and (\ref{angles2}).}
\label{fig_bloc_depth}
\end{figure}

The computed flow field in the gas phase allows one to immediately deduce  $\beta_{\parallel,\perp}$. Fig.~\ref{fig_bloc_depth} shows $\beta_{\parallel,\perp}$ for trapezoidal grooves of  a fixed $\alpha=3\pi/32$ and several depths $e^*/\delta$. We see that for shallow grooves ($e^*/\delta \ll 1$)  the $\beta_{\parallel,\perp}$
profiles can be approximated by trapezoids with the central region of a constant slip $\beta_{\parallel} \simeq e/\delta$ and $\beta_{\perp} \simeq e/4 \delta$ given by
Eqs.(\ref{shallow}), and linear regions near edges with local slip coefficients
described by Eqs.(\ref{angles}) and (\ref{angles2}). For relatively deep grooves, $e^*/\delta = O(1)$, $\beta_{\parallel,\perp}$ profiles practically converge into
single curves and are no longer dependent on $e^*/\delta$. We also see that the crossover between the regimes of shallow and deep grooves takes place at intermediate $e^*/\delta$, where $\beta_{\parallel,\perp}$ are controlled both by $e^*/\delta$ and $\alpha$. To illustrate this we have included in Fig.~\ref{fig_bloc_depth} the $\beta_{\parallel,\perp}$ curves computed for arc-shaped grooves of the same $\alpha=3\pi/32$, which implies their $e^{\ast}/\delta \simeq 0.37$. Remarkably, the $\beta_{\parallel,\perp}$ profiles are very close to computed for trapezoidal grooves of $e^*/\delta = 0.35$. Some small discrepancies in the shapes of $\beta_{\parallel,\perp}$ obtained for two types of textures indicate that in the crossover regime the shape of the grooves contributes to a slip length profile, but this should be seen as a second-order correction only.

\begin{figure}[tb]
\includegraphics[width=0.75\columnwidth]{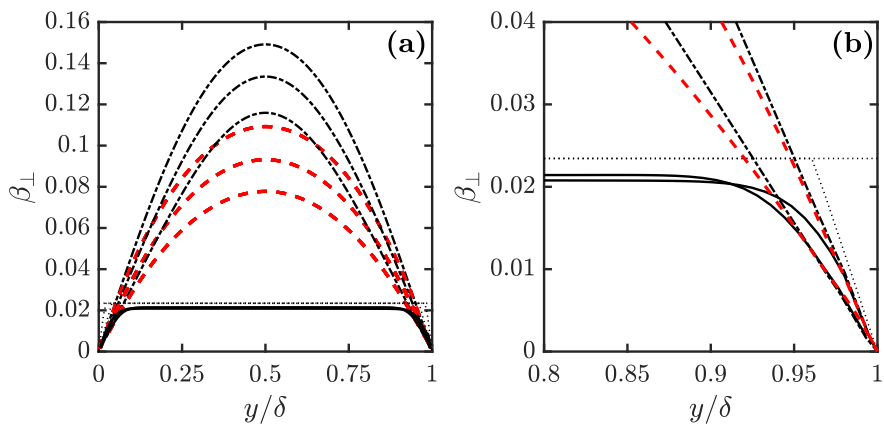}
\caption{(a)~Transverse slip coefficient profiles for deep trapezoidal grooves of $e^*/\delta=0.97$ (dash-dotted curves), shallow trapezoidal grooves of $e^*/\delta=0.094$ (solid curves).  From top to bottom $\alpha=0,~\pi/16$, and $\pi/8$. Dotted line represents prediction of Eq.~(\ref{shallow}) at $e^*/\delta=0.094$. Dashed curves plot the results for arc-shaped grooves of the same values of $\alpha$, which implies $e^{\ast}/\delta$ is varying from 0.5 down to 0.32. (b)~The data sets for $\alpha=0$ reproduced in a larger scale in the vicinity of the edge, $y/\delta = 1$,  together with  $\beta_{\perp}$ curves computed at $\alpha=3\pi/32$. }
\label{fig_bloc}
\end{figure}

Now we focus on the role of $\alpha$ and study first only deep, $e^*/\delta=0.97$, and shallow, $e^*/\delta=0.094$, trapezoidal grooves  The computed $\beta_{\perp}$ are displayed in Fig.~\ref{fig_bloc}(a). One can see that in the case of shallow textures $\beta_{\perp}$ generally does not depend on $\alpha$  and is consistent with Eq.(\ref{shallow}), but note that there are deviations from Eq.(\ref{shallow}) in the vicinity of the edges. For deep trapezoids it decreases with $\alpha$. These trends fairly agree with the predictions of Eqs.(\ref{angles}) and Eqs.(\ref{angles_prim}). We can now compare $\beta_{\perp}$ for deep trapezoidal grooves with that of arc-shaped grooves with the same bevel angle. The results of calculations are included to Fig.~\ref{fig_bloc}(a). The data show that at the central part of the gas sector the arc-shaped grooves induce smaller $\beta_{\perp}$ than trapezoidal ones, which simply reflects the fact that they are not deep enough. Indeed, with our values of $\alpha$, their maximal depth $e^{\ast}/\delta$ varies from $0.5$ down to $0.32$ indicating the crossover regime. However, in the vicinity of the groove edges $\beta_{\perp}$ for both textures appears to be the same. To examine this more closely, the data  obtained with two values of $\alpha$ in the edge region is plotted in Fig.~\ref{fig_bloc}(b), and we see that in the vicinity of the point where three phases meet $\beta_{\perp}$ computed for different grooves indeed coincide. Altogether these results do confirm that $\beta'_{\perp}$ is determined by $\alpha$ only.

\begin{figure}[tb]
\includegraphics[width=0.75\columnwidth]{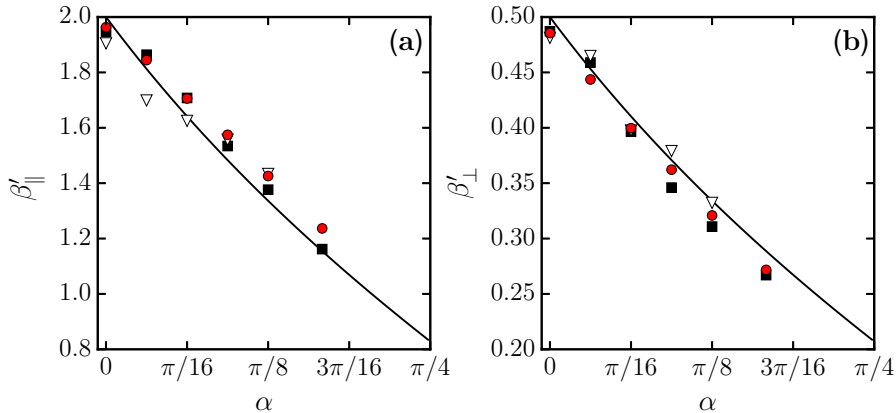}
\caption{Computed (a)~$\beta^\prime_{\parallel}$ and (b)~$\beta^\prime_{\perp}$ plotted as a function of $\alpha$ (symbols). Squares and triangles indicate trapezoidal grooves of $e^*/\delta=0.97$ and $0.094$, circles plot the results for arc-shaped grooves. Calculations made with Eq.(\ref{angles_prim}) are shown by solid curves.   }
\label{fig_angles}
\end{figure}

Fig.~\ref{fig_angles} includes $\beta'_{\parallel,\perp}$ curves computed for  deep, $e^*/\delta=0.97$, and shallow, $e^*/\delta=0.094$, trapezoidal grooves and arc-shaped grooves. The calculations are made using several $\alpha$ in the range from $0$ to $\pi/6$, which implies that $e^{\ast}/\delta$ varies from 0.5 down to 0.29.  The general conclusion from this plot is that $\beta'_{\parallel,\perp}$ does not depend on the groove shape and depth being a function of $\alpha$ only, so that our results fully verify Eq.(\ref{angles_prim}).

\begin{figure}[tb]
\includegraphics[width=0.75\columnwidth]{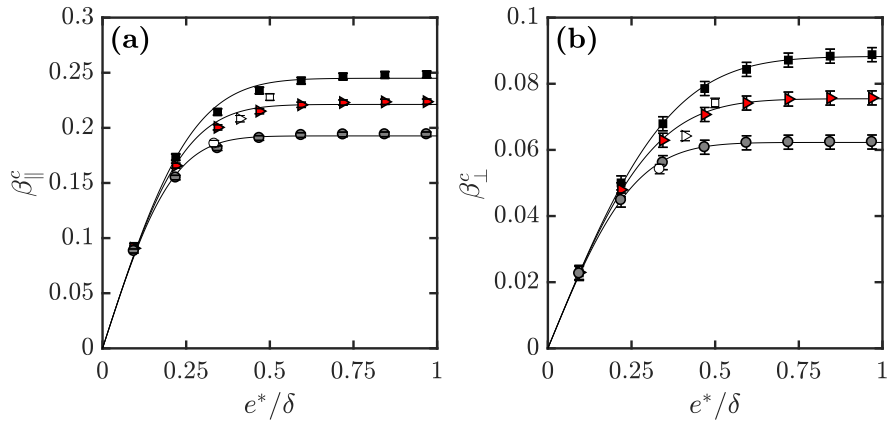}
\caption{Apparent slip coefficients, $\beta_{\parallel,\perp}^c$, as a function of $e^*/\delta$ computed for trapezoidal  grooves (filled symbols). From top to bottom $\alpha=0$, $\pi/16$, and $\pi/8$. Error bars were determined from the deviation among repeated calculations made at different $\phi$. When error bars are absent, uncertainties are
smaller than the symbols. Solid curves are plotted to guide the eye. Open symbols correspond to $\beta_{\parallel,\perp}^c$ of arc-shaped grooves of the same $\alpha$.}
\label{fig_bapp}
\end{figure}

We now turn to the effective slip lengths, $b_{\mathrm{eff}}^{\|,\perp}$, and will try to understand if such a two-phase problem can indeed be reduced to a one-phase problem, and whether a SH surface can be modeled as a flat one with patterns of stripes with piecewise constant apparent local slip lengths $b^{c}_{\parallel,\perp}$. We first fit our numerical data for $b_{\mathrm{eff}}^{\|,\perp}$ to the known formulae~\cite{belyaev.av:2010a}:
\begin{equation}
\begin{array}{ll}
b_\mathrm{eff}^{\|}\simeq\dfrac{L}{\pi}\dfrac{\ln\left[\sec\left(\frac{\pi\phi}{2}\right)\right]}{1+\dfrac{L}{\pi b^c_{\parallel}}\ln\left[\sec\left(\frac{\pi\phi}{2}\right)+\tan\left(\frac{\pi\phi}{2}\right)\right]},
\\\label{belyaev}
b_\mathrm{eff}^{\perp}\simeq\dfrac{L}{2\pi}\dfrac{\ln\left[\sec\left(\frac{\pi\phi}{2}\right)\right]}{1+\dfrac{L}{2\pi b^c_{\perp}}\ln\left[\sec\left(\frac{\pi\phi}{2}\right)+\tan\left(\frac{\pi\phi}{2}\right)\right],}
\end{array}\end{equation}
taking $b^{c}_{\parallel,\perp}$ as fitting parameters, and then deduce $\beta^c_{\parallel,\perp}$, which can be defined as~\cite{nizkaya2014gas}
\begin{equation}
\beta^c_{\parallel,\perp} = \dfrac{\mu_g }{\mu} \frac{b^{c}_{\parallel,\perp}}{\delta}
\label{newb2}
\end{equation}%
We stress that with such a definition the eigenvalues of apparent slip coefficients, $\beta^c_{\parallel,\perp},$ of trapezoidal grooves depend only on $\alpha$ and
$e^*/\delta$, but not on $y/\delta$. The curves for $\beta ^c_{\parallel,\perp},$ computed for several $\alpha$ are plotted Fig.~\ref{fig_bapp}. For all $\alpha$ these functions saturate already at $e^*/\delta \geq 1,$ thereby imposing constraints on the attainable $b_{\mathrm{eff}}^{\|,\perp}$. Also included in Fig.~\ref{fig_bapp} are $\beta ^c_{\parallel,\perp},$ for arc-shaped texture of the same $\alpha$. In this case $e(y, \alpha)/\delta$ is nonuniform throughout the cross-section, and we make no attempt to calculate its average or effective value at a given $\alpha$. Instead, Fig.~\ref{fig_bapp} is intended to indicate the range of $\beta ^c_{\parallel,\perp},$ that is expected for arc-shaped grooves, so that in this case we simply plot data as a function of $e^{\ast}/\delta$. One can see that data for arc-shaped  grooves nearly coincide with results for trapezoidal grooves at $\alpha = \pi/8$,  but at smaller angles there is
some discrepancy, especially when $\alpha=0$ and for a transverse case. The discrepancy is always in the direction of smaller $\beta ^c_{\parallel,\perp},$ than found for trapezoidal grooves of $e=e^{\ast}$, indicating that effects of the groove shape become discernible at intermediate depths. We stress, however, that both for trapezoidal and arc-shaped grooves the values of $\beta ^c_{\parallel,\perp},$ are close and show the same trends. So our results for two types of grooves provide a good sense of the possible local slip of 1D texture of any shape.

 These results  can be used to predict upper attainable local and effective slip lengths of complex grooves if their bevel angle and maximal depth are known. Indeed, Fig.~\ref{fig_bapp} allows one to immediately evaluate $\beta ^c_{\parallel,\perp},$ and the apparent local slip lengths can be found by using Eq.(\ref{newb2}). Once they are known, the eigenvalues of the effective length tensor can be calculated analytically by using  Eqs.(\ref{belyaev}).

\section{Conclusion}

We have analyzed liquid slippage at gas areas of 1D SH surfaces and developed an asymptotic theory, which led to
explicit expressions for the longitudinal and transverse slip lengths near the edge of the groove, i.e. the point, where three phases meet. The theory predicts that the local slip lengths in the vicinity of this point always increase linearly with the slope determined solely by a bevel angle of grooves. We have also shown that at a given viscosity contrast the eigenvalues of local slip tensor of strongly slipping deep grooves are fully determined by their width and the value of their bevel angle, but not by their depth as sometimes invoked for explaining the extreme local slip. Thus, it is not necessary to deal with very deep grooves to get a largest possible local slip at the gas areas. However, the eigenvalues of a local slip tensor of weakly slipping shallow grooves are not really sensitive to the bevel angle and are determined mostly by their depth.

Altogether, our study shows that for a given width, SH grooves with beveled edges are less efficient than rectangular ones for drag reduction purposes. However, the use of grooves with beveled edges appears as a good compromise between the positive effect of their stability against bending and a moderate reduction of the slippage effect due to a bevel angle. A very large local slip length can be induced by using wide grooves with beveled edges, which would be often impossible for wide rectangular grooves due to their bending instability.

To check the validity of our theory, we have proposed an approach to solve numerically the two-phase hydrodynamic problem by considering separately flows in liquid and gas phases, which can in turn be done by using different techniques. Our method significantly facilitates and accelerates calculations compared to classical two-phase numerical schemes. Generally, the numerical results have fully confirmed the theory for limiting cases. They have also clarified that at intermediate depths a particular shape of 1D SH surface modifies the local slip profiles, but only slightly.

Our strategy and computational approach can be extended to a situation of a sufficiently curved meniscus. Another fruitful direction could be to consider more complex 2D textures, which include various pillars and holes. Thus, they may guide the design of textured surfaces with superlubricating potential in microfluidic devices, tribology, and more. It would be also interesting to revisit recent analysis of an effective slip of SH surfaces and its various implications since our results suggest that instead of a piecewise constant slip length at the gas areas, a local tensorial slip should be employed to obtain more accurate eigenvalues of the effective slip length tensor.  Finally, we mention that our approach can be immediately applied to compute local slip lengths of grooves filled by immiscible liquid of low or high viscosity.


\begin{acknowledgments}
This research was partly supported by the the Russian
Foundation for Basic Research (grants No. 15-01-03069).
\end{acknowledgments}

\appendix
\section{Local slip lengths near the edge of the groove.}
\label{sec_app_edge}

 \begin{figure}[tb]
\includegraphics[width=0.375\columnwidth]{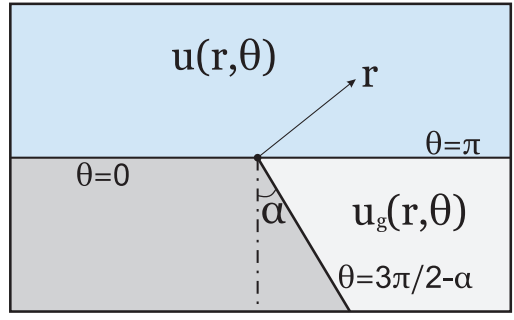}
\caption{Polar coordinates in the vicinity of the groove edge.}
\label{fig_appendix}
\end{figure}

Here we obtain the solutions of the Stokes equations and explicit formulas for eigenvalues of the local slip length in the vicinity of the groove edge $\left( y,z\right)=\left(0,0\right) $. Following ~\cite{wang2003,Zhou_etal:2013} we use polar coordinates $(r,\theta)$, so that $y/\delta=-r\cos \theta ,$ $
z/\delta=r\sin \theta $ (see Fig.\ref{fig_appendix}), and focus on the case $r\ll 1$. We will show that in this situation the eigenvalues of the local slip length, $b_{\parallel,\perp}$ augment linearly with $r$, are proportional to the viscosity ratio when it is large, $\mu/\mu_g\gg 1$, and decrease with $\alpha$.

 In the case of longitudinal grooves the velocity has only one component $\mathbf u=(u,0,0)$ and the Stokes equations reduce to the Laplace equation, $\Delta u=0$. The general solution implies a power-law dependence of
velocities on the distance:
\begin{eqnarray}
u_{l,g}&=&r^{\lambda_{\parallel} }\left[ a_{l,g}\sin \left( \lambda_{\parallel} \theta \right)
+c_{l,g}\cos \left( \lambda_{\parallel} \theta \right) \right],
\label{u_x}
\end{eqnarray}
where unknown exponent $\lambda_{\parallel}$ and coefficients $a_l,a_g,c_l,c_g$ can be found by imposing boundary conditions. These are defined in the usual way. Namely, we use the no-slip boundary conditions at the solid walls
  \begin{equation}\label{noslip}
    u_l(r,0)=0, \quad u_g(r,\theta_w)=0
  \end{equation}
  where $\theta_w=3\pi /2-\alpha$, and at the liquid-gas interface, $\theta =\pi$, we impose
\begin{equation}
 u_{l}=u_{g},\quad
\mu \partial _{\theta}u_{l}=\mu _{g}\partial _{\theta}u_{g}.%
 \label{bc_i}
\end{equation}
Applying these boundary conditions, we obtain a linear homogeneous system
\begin{equation}
c_l=0,
\label{bc_l}
\end{equation}
\begin{equation}
a_{g}\sin \left( \lambda_{\parallel} \theta_w \right)
+c_{g}\cos \left( \lambda_{\parallel} \theta_w \right) =0,
\label{bc_g}
\end{equation}
\begin{equation}
(a_l-a_{g})\sin \left( \lambda_{\parallel} \pi \right)+
(c_l-c_{g})\cos \left( \lambda_{\parallel} \pi \right)=0,
\label{bc_is}
\end{equation}
\begin{equation}
(\mu a_l-\mu _{g}a_{g})\cos \left( \lambda_{\parallel} \pi \right)-
(\mu c_l-\mu _{g}c_{g})\sin \left( \lambda_{\parallel} \pi \right)=0,
\label{bc_iv}
\end{equation}
which allows us to derive an equation for $\lambda_{\parallel}$ by eliminating $a_{l},\ a_{g},\ c_{g}$,
\begin{equation}
\tan \left[ \lambda_{\parallel} \left(\frac{3 \pi}{2}-\alpha\right)\right] =\frac{\left( \mu-\mu_g\right)
\tan \left( \lambda_{\parallel} \pi \right) }{\mu+\mu_g\tan^2\left( \lambda_{\parallel} \pi
\right) }.  \label{lam1}
\end{equation}%
This result implies that the value of $\lambda_{\parallel}$ is determined by $\mu/\mu_g$ and $\alpha$.


By expanding (\ref{lam1}) into a Taylor series in $\mu_g/\mu\ll 1$ we get
\begin{equation}
\lambda_{\parallel} =\frac{1}{2}-\frac{\mu_g}{\mu}\dfrac{\tan(\pi/4+\alpha/2)}{\pi}+O\left(\dfrac{\mu_g^2}{\mu^2}%
\right) .
\label{exp} \end{equation}

 The longitudinal slip length profile near the edge of the groove can be then calculated as
 \begin{equation}
 b_{\parallel} = \dfrac{r \delta u_{x}(r, \pi)}{\partial _\theta u_x(r, \pi )}=-\dfrac{r \delta \tan \left( \lambda_{\parallel}  \pi \right) }{\lambda_{\parallel}  }.
 \label{bx}
\end{equation}
We note that the last equality has been obtained by using Eq.(\ref{bc_l}). By using Eqs.(\ref{exp}) and (\ref{bx}) we can then derive for
$\mu/\mu_g\gg 1$
\begin{equation}
b_{\parallel} \simeq \frac{\mu }{\mu_g}\frac{2y}{\tan(\pi/4+\alpha/2)}.
\label{bxAsympt}
\end{equation}%

In the case of transverse grooves, the solution of Stokes equations can be expressed in terms of a streamfunction $\psi $ which satisfies the biharmonic equation, $\Delta ^{2}\psi =0$. A general solution  in the liquid phase has the following form~\cite{wang2003}:
\begin{equation}
\begin{array}{ll}
\psi_l (r,\theta ) & =r^{\lambda_{\perp} }\left[ a_l\sin (\lambda_{\perp} \theta )+g_l\sin
((\lambda_{\perp} -2)\theta )\right.  \\
& +\left. c_l\cos (\lambda_{\perp} \theta )+h_l\cos ((\lambda_{\perp} -2)\theta )\right],
\end{array}
\label{tr1}
\end{equation}%
and the radial and the angular components of the velocity are
\begin{equation}
u_{lr}(r,\theta )=\dfrac{\partial _{\theta }\psi }{r},\quad u_{l\theta }(r,\theta
)=-\partial _{r}\psi .  \label{tr2}
\end{equation}%
The same equations are valid for a streamfunction $\psi _{g}$ and velocities $u_{gr}$,  $u_{g\theta }$, in the gas phase, with coefficients $a_g, g_g, c_g$ and $h_g$ in Eq.(\ref{tr1}). To find the eight unknown coefficients and $\lambda_{\perp}$ we again apply the boundary conditions at the solid walls and at the liquid-gas interface. The no-slip conditions at the wall, $u_{r}=u_{\theta }=0$ at $\theta=0$ and $\theta=\theta_w$ allow us to formulate four equations of the system:
\begin{equation}
c_l+h_l=0, \label{nst}
\end{equation}%
\begin{equation}
\lambda_{\perp} a_l+(\lambda_{\perp} -2)g_l=0, \label{nsr}
\end{equation}%
\begin{equation}
a_g\sin (\lambda_{\perp} \theta_w )+g_g\sin ((\lambda_{\perp} -2)\theta_w )+c_g\cos (\lambda_{\perp} \theta_w
)+h_g\cos ((\lambda_{\perp} -2)\theta_w )=0,
\end{equation}%
\begin{equation}
\lambda_{\perp} a_g\cos (\lambda_{\perp} \theta_w )+(\lambda_{\perp} -2)g_g\cos ((\lambda_{\perp} -2)\theta_w
)-\lambda_{\perp} c_g\sin (\lambda_{\perp} \theta_w )-(\lambda_{\perp} -2)h_g\sin ((\lambda_{\perp} -2)\theta_w )=0.
\end{equation}%

The conditions at the interface, $\theta=\pi$, namely of  the impermeability, $u_{l\theta}=u_{g\theta}=0$, and of the continuity of the tangent velocity, $u_{lr}=u_{gr}$, and the shear stress,
$\mu \partial _{\theta}u_{lr}=\mu _{g}\partial _{\theta}u_{gr}$, give
\begin{equation}
\left( a_l+g_l\right) \sin (\lambda_{\perp}
\pi )+\left( c_l+h_l\right) \cos (\lambda_{\perp}  \pi ) =0,  \label{vnl}
\end{equation}
\begin{equation}
\left( a_g+g_g\right) \sin (\lambda_{\perp}
\pi )+\left( c_g+h_g\right) \cos (\lambda_{\perp}  \pi ) =0,  \label{vnr}
\end{equation}
\begin{equation}
\left[ (a_l-a_g)\lambda_{\perp}  +(g_l-g_g)(\lambda_{\perp}  -2)\right]
\cos (\lambda_{\perp}  \pi )-\left[ (c_l-c_g)\lambda_{\perp}  +(h_l-h_g)(\lambda_{\perp}  -2)\right] \sin (\lambda_{\perp}  \pi
)=0,  \label{ur}
\end{equation}
\begin{equation}
\left[ (\mu a_l-\mu _{g}a_g)\lambda_{\perp} ^{2}+(\mu g_l-\mu _{g}g_g)(\lambda_{\perp}  -2)^{2}\right] \sin (\lambda_{\perp}  \pi
)+\left[ (\mu c_l-\mu _{g}c_g)\lambda_{\perp} ^{2}+(\mu h_l-\mu _{g}h_g)(\lambda_{\perp} -2)^{2}\right] \cos (\lambda_{\perp}  \pi )=0.
\label{urt}
\end{equation}%

 The solution of a system of Eqs.(\ref{nst})-(\ref{urt}) cannot be reduced to a single equation for $\lambda_{\perp}$, as it has been done in the longitudinal case (and led to $\lambda_{\parallel}$ described by Eq.(\ref{exp})). However, it is possible to obtain an asymptotic solutions when $\mu/\mu_g \gg 1$:
\begin{equation}
 \lambda_{\perp} =\frac{3}{2}-\frac{\mu_g}{\mu}\dfrac{2\tan(\pi/4+\alpha/2)}{\pi}+O\left(\dfrac{\mu_g^2}{\mu^2}%
\right).
\end{equation}

 The local transverse slip length can be calculated  as
 \begin{equation}
 b_{\perp} = \dfrac{r\delta u_{lr}(r, \pi)}{\partial _\theta u_{lr}(r, \pi )}=\dfrac{r \delta \tan \left( \lambda_{\perp}  \pi \right) }{2(\lambda_{\perp}-1)  },
 \label{by}
\end{equation}
where the last equality is obtained by using Eqs.(\ref{nst}) and (\ref{nsr}).

This immediately gives
\begin{equation}
b_{\perp}\simeq \dfrac{\mu }{\mu _{g}}\dfrac{y}{2\tan(\pi/4+\alpha/2)}=\dfrac{b_{\parallel}}{4}.
\label{byAsympt}
\end{equation}

\bibliography{viscosity}

\begin{thebibliography}{47}
\bibitem [{not()}]{note3}%
  \BibitemOpen
  \href@noop {} {}\bibinfo {note} {As a side note we should like to mention
  that Eqs. (\ref{bgas}), (\ref{bpadef}) and (\ref{bpedef}) suggest that
  $\delta \beta_{\parallel,\perp}$ can be interpreted as eigenvalues of the
  local slip lengths on the gas side of the interface.}\BibitemShut {Stop}%
\end{thebibliography}

\end{document}